# SEARCHING FOR DARK MATTER AT COLLIDERS

**Francois Richard[1,*], Giorgio Arcadi[2,§] and Yann Mambrini[2,¤]**

*1 Laboratoire de l'Accélérateur Linéaire, IN2P3/CNRS
and Université Paris-Sud 11 Centre Scientifique d'Orsay,
B. P. 34, F-91898 Orsay Cedex, France and
2 Laboratoire de Physique Théorique Université Paris-Sud, F-91405 Orsay, France*

*Abstract*

Dark Matter (DM) detection prospects at future e+e- colliders are reviewed under the assumption that DM particles are fermions of the Majorana or Dirac type. Although the discussion is quite general, one will keep in mind the recently proposed candidate based on an excess of energetic photons observed in the center of our Galaxy with the Fermi-LAT satellite. In the first part one assumes that DM particles couple to vector bosons, either the SM Z or a Z'. Taking into account the strong constraints set by direct searches, in particular the LUX experiment, one assumes that DM is made of Majorana fermions. While this solution accommodates LUX limits, it appears incompatible with the Indirect evidence from Fermi-LAT unless one invokes the presence of Sommerfeld forces to enhance the annihilation rate at present temperatures. At future colliders, the most sensitive measurement comes from the Z invisible width and allows, at best, to reach a mass limit $m_\chi > 35$ GeV. If one assume that DM couples to a Z', it becomes possible to allow Dirac DM particles, provided that this Z' only couples axially to SM fermions. To satisfy the cosmological constraints, this Z' should have a mass below 1 TeV and tends to decay invisibly in more than 90% of the cases. With reduced couplings to standard fermions, it remains undetected at LHC. Using radiative return events e+e-→XX+γ, ISR, one could observe a spectacular signal at a TeV e+e- collider. This result relies on the ability of using highly polarized beams to eliminate a large part of the W exchange background. Prospects of discovery at LHC using mono-jets are also discussed and appear promising. In the second part, one assumes that DM particles annihilate through Higgs particles, either the SM boson h or MSSM type bosons called H, A. A promising scenario emerges, where one has e+e-→HA, with H decaying into hh, while A decays invisibly in most of the cases. In such a situation, with well defined initial energy momentum and with adequate detectors to reconstruct the 2h state, one can observe a clear A signal using the missing mass technique.

*richard@lal.in2p3.fr
§giorgio.arcadi@th.u-psud.fr
¤Yann.mambrini@th.u-psud.fr

# I Introduction

Search for dark matter is of prime importance for our understanding of the universe. This goal is pursued using a wide variety of approaches, given the very large spectrum of interpretations predicting particles with a mass range between μev, multi TeV and even beyond, from axion to wimpzilla. Several direct detection (DD) searches provide signals originating from underground experiments but without converging evidence. There are also several indirect detection (ID) hints based on photons coming from the center of the galaxy (3.5 kev and 130 GeV lines, photon excess in the GeV range) or from positron excess. No consistent picture emerges so far from both types of searches.

The DD remaining candidates, which tend to cluster at low masses, 10 to 20 GeV, seem contradicted by recent results from superCDMS. This is also true for LUX and Xe100 experiments which are reaching a very high level of sensitivity which already covers a large set of predictions assuming a spin independent scattering of DM with nuclei.

The signals from ID can be attributed to classical sources, like pulsars or supernovae remnants, for the positrons and for the Fermi LAT photons coming from the center of our Galaxy.

Collider searches are therefore the necessary complement for a safe conclusion on this essential investigation. Here we will focus on the prospects offered by future e+e- colliders, in particular the International Linear Collider, ILC, with polarized beams, keeping in mind the genuine wimp interpretation of the Fermi LAT candidate and the constraints from the LUX, the invisible Z width from LEP1 and the invisible H width from LHC.

We will accordingly pick up 2 Standard Model (SM) type portals where fermionic (Dirac or Majorana) DM annihilation takes place through Z and the SM Higgs boson. This approach will be extended to two generic BSM models: one assuming a Z' portal, the other assuming a non minimal Higgs sector. No specific assumption will be made about the origin of these fermionic DM particles, of the type MSSM or NMSSM, which allows to freely vary their couplings to vector and scalar bosons.

# II The galactic center photon excess

The gamma-ray excess reported in [1] seems relevant for accelerator searches since it could be interpreted as the annihilation of massive dark matter particles, possibly into b jets (~35 GeV mass) or democratically into SM fermions (~25 GeV mass). Quoting [1] 'the signal is observed to extend to at least ~$10^0$ from the Galactic Center (GC), disfavoring the possibility that this emission originates from millisecond pulsars'.

More recently [2] a thorough analysis of this Fermi photon GeV excess has been studied. Assuming the interpretation of reference [1], the estimated DM mass is higher, 49±6 GeV.

Various interpretations of this annihilation process can be provided with a minimum extension of the SM, meaning that one can try to reproduce correctly the annihilation cross section claimed by [1] by assuming that DM couples to SM particles like the Z or the Higgs boson. In doing so, one can take into account existing accelerator limits on invisible decay of these particles which, as will be seen, can provide essential constraints.

It will also be necessary to cope with the strong limits provided by the LUX experiment for spin independent, SI, interactions which reaches its full sensitivity in the mass region claimed for the Fermi-LAT signal. Recall however that the SI cross section limits assume coherent recoil of the nucleus caused by the DM scattering. For a heavy nuclear target, the coherent scattering increases the cross-section by the square of the Atomic Number. This is not the case for spin dependent, SD, cross-section which occur through the axial vector coupling to the spin content of the nucleus, meaning that the cross section limits are about 4 orders of magnitude weaker than for SI. Recall that a spin coupling is the only possibility if one assumes that the DM fermion X is a Majorana particle. The LUX limits can therefore be relaxed by assuming that Z/Z' couples axially to DM.

With this choice however one finds that the annihilation cross section through a Z boson depends on v², v being the velocity of DM particles which gets suppressed at present temperatures and is therefore usually discarded as incompatible with the photon excess from the galactic center. We will however keep open this interpretation by assuming that low velocity suppression can be compensated by a Sommerfeld enhancement due to a new interaction as will be further discussed in III.2. This problem can be avoided for a Z' where one is free to assume an axial coupling for SM fermions.

For what concerns scalar mediators, of Higgs type, one can also assume an axial coupling $\bar{X}\gamma_5 X$ to DM but this solution is excluded by present LHC limits on invisible Higgs decays given that this particle should decay predominantly into DM if we want to explain the observed excess (disregarding a Sommerfeld enhancement). Extended Higgs models provide viable solutions, in particular using the pseudo-scalar A component present in the two Higgs doublet scheme, as discussed in section VIII.

| DM | Mediator | Interactions | Direct | LHC |
|---|---|---|---|---|
| Majorana | Z' | $\bar{X}\gamma^\mu\gamma^5 X$ , $\bar{f}\gamma^\mu\gamma^5 f$ | Yes | Yes |
| Dirac | Z' | $\bar{X}\gamma^\mu X$ , $\bar{f}\gamma^\mu\gamma^5 f$ | No | Yes |
| Majorana | A | $\bar{X}\gamma^5 X$ , $\bar{f}\gamma^5 f$ | No | Yes |

Above table, extracted from [3], summarizes the prospects of confirmation of the GC photon excess for the type of couplings envisaged in this note. It is remarkable that that LHC can cover scenarios which are not accessible by DD and our task will be to analyze in which ways an e+e- collider can complement LHC for a better exploration of this type of signal.

# III Z portal

To reproduce the amount of primordial DM, one assumes [3,4] an annihilation of DM Majorana fermions XX→Z→ffbar with axial coupling without co-annihilation processes. At thermal freeze-out, with <v>~0.3, the cross section has to satisfy the canonical value needed to provide the observed amount of DM in our universe: <σv>~3 10$^{-26}$cm$^3$/s.

### III.1 Thermal freeze-out

The couplings are defined by:

$$\mathcal{L}_{int} \supset \left[ a\bar{X}\gamma^\mu \left( g_V^X + g_A^X \gamma^5 \right) X \right] Z_\mu$$ where a=1 for Dirac and a=1/2 for Majorana (with $g_V^X$ =0)

Neglecting the fermion masses, at decoupling [3,5] one has:

$$\sigma v_{FO} = \sum_f n_{cf} (|g_V^f|^2 + |g_A^f|^2) \frac{|g_A^X|^2 \langle v^2 \rangle s}{12\pi \left[ (s-m_Z^2)^2 + (m_Z\Gamma_Z)^2 \right]}$$

with $s = 4m_X^2$ $g_V^f = \frac{g}{2c_W}(I_3 - 2Qs_W^2)$ $g_A^f = \frac{g}{2c_W}I_3$. Summing up on all fermion final states gives approximately $\Sigma_f \sim 1$. Note that this formula is valid both for Dirac and for Majorana fermions (see [3] appendix 4). Here one takes an average $\langle v^2 \rangle$ =0.24. This approximation is well justified except near the resonance, where, as pointed out in [6], on needs a more precise calculation.

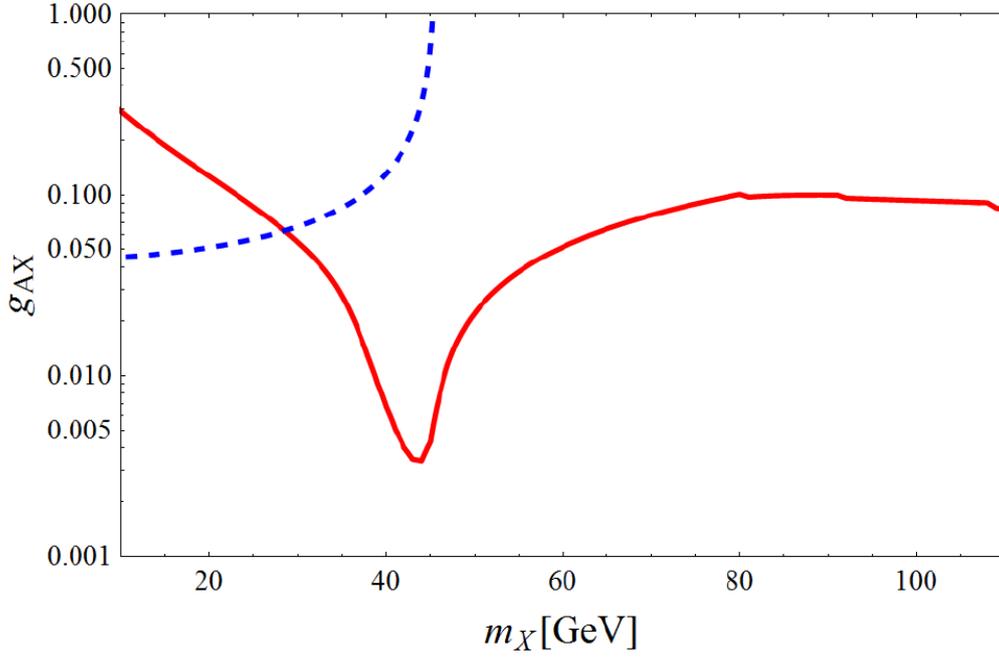

Figure 1: Predicted axial coupling of Z to Majorana DM fermions versus their mass. The blue dashed curve comes from the Z invisible width limit from LEP1.

In figure 1, the curve from [7] shows the dependence of the axial coupling $g_A^X$ with respect to $m_X$. This curve, which relies on the exact expression of the annihilation cross section, without performing the velocity expansion, differs appreciably, up to a factor 2 at resonance, from our naïve formula. This detail is of importance, recalling that in reference [2] gives a DM mass estimate of 49±6 GeV.

### III.2 Annihilation signal from the galactic center

After decoupling, our universe cools down and, at present, the velocity of DM is ~300km/s, that is <v>=0.001. This means that the annihilation cross section previously computed becomes completely negligible and therefore unable to explain the indirect signal observed by Fermi-LAT. Note however that our calculation has neglected the fermion masses which is not legitimate for the b quark when <v>→0.

Neglecting the $\langle v^2 \rangle$ term and normalizing to the previous cross section:

$$\frac{\sigma v_{GC}}{\sigma v_{FO}} = 3 BR(Z \to b\bar{b}) \frac{m_b^2}{s\langle v^2 \rangle} \left[ 1 - (s/m_Z^2) \right]^2$$ The final state is dominantly made of b jet pairs

which satisfies [1] but with an annihilation cross section ~1000 smaller than at freeze-out.

➔ This scenario has been rejected by [3] since it appears inconsistent to explain both the amount of DM at freeze-out and the large signal observed by Fermi-LAT.

At this point one may recall that the detailed distribution of DM at the galaxy level does not match the ΛCDM model which assumes non interacting DM particles. It has therefore been proposed to invoke a *Sommerfeld type mechanism* with the exchange of a light, ~10 MeV, particle which could considerably, by a few 100, enhance the rate at very low velocity [8]. This mechanism would therefore save a Z exchange interpretation of the Fermi-LAT indication.

**III.3 The Z invisible width and the ISR measurement.**

The Z invisible width has been very precisely measured at LEP1 and can be modified if there is a substantial decay of Z into X Majorana fermions. One has:

$$\Gamma(Z \to XX) = \frac{|g_A^X|^2 v^3 m_Z}{24\pi} \text{ where } v = \sqrt{1 - \frac{4m_X^2}{m_Z^2}}$$

The LEP1 upper limit for the BSM invisible width being 2 MeV, one can exclude solutions with $m_X$<27 GeV, which is compatible with the interpretation given in [1]. For a Dirac fermion, with an axial coupling, one has $m_X$<29 GeV.

At future e+e- colliders, a factor of ~2 in precision appears feasible taking into account the dominant contribution due to luminosity accuracy at 0.1%. This gives $m_X$<28.5 GeV for a Majorana fermion.

An alternative method uses radiative return to the Z peak by running at a circular collider [9] above this peak, at maximum integrated luminosity. One is limited by systematical errors but, as argued by [9], using the leptonic modes for normalization, one can remove most uncertainties. This approach could achieve up to an order of magnitude accuracy improvement. Even then, the invisible Z width method can only cover masses up to 35 GeV.

**III.4 Discussion.**

In the Z portal scenario, LEP results can exclude a Majorana fermion with mass below $m_X$=27 GeV, insufficient to interpret/exclude the GC photon excess reported in [1]. Future e+e- colliders will reach at best $m_X$=35 GeV. For what concerns LHC, given the predicted branching ratio of Z into DM, no observable signal can be seen by the monojet search above the large background due to Z decays into neutrinos.

# IV A Z' portal ?

As we shall see, a Z' exchange scenario [10] offers many opportunities. In the same way that one does not specify the theoretical embedding of the DM fermions, one can remain vague about the origin of such a Z' and simply parameterize its couplings to ordinary matter with a scale factor κ² with respect to a sequential Z' (i.e. a heavy Z with the SM couplings).
Also one can solve the inconsistency on the annihilation cross section at very low velocity, already discussed in III.2, if one assumes that Z' only couples axially to matter and vectorially to Dirac DM fermion. In this case LUX constraints are fully avoided since there is a kinematical suppression [3].

This mechanism is velocity independent and therefore maintains a large annihilation cross section even when v goes to zero. In what follows one will therefore envisage two scenarios:

- Scenario 1 a Z' having an axial coupling to ordinary matter and a vector coupling to a Dirac X
- Scenario 2 a Z' coupling as a Z to matter with axial coupling to a Majorana X

In both cases one allows for a suppression factor K² with respect to Z SM couplings.

In this type of scenario, as discussed below, the Z' will decay *mostly invisible*, which requires using the ISR technique in e+e- . In the next section the modus operandi for this technique is described for what concerns the ILC set up.

### IV.1 The ISR approach at ILC

To cover a heavy invisible Z' scenario, one needs to operate at high energies and use initial state radiation (ISR) at angle. Above the Z pole, the main background comes from e+e→$\nu_e\nu_e\gamma$ with W exchange (see diagrams below). This process is only sensitive to left handed electrons and therefore can be efficiently removed using right handed polarization for electrons which can be provided by ILC [11].

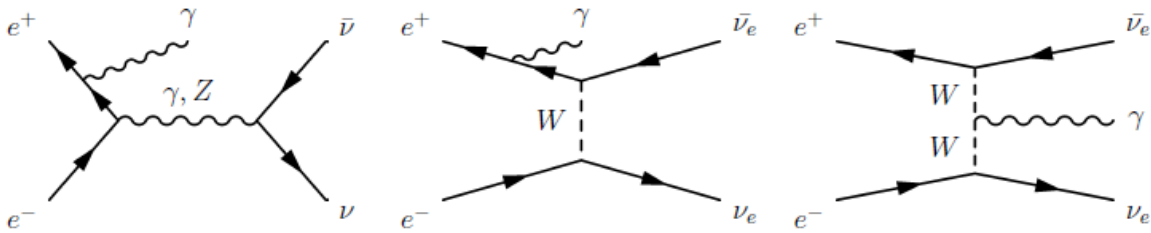

One can also assume an improved polarization [12] with respect to the base line ILC parameters: Pe-=90% and Pe+=-60% instead of Pe-=80% and Pe+=-30%. The corresponding suppression of the W exchange process improves by a factor 4. To understand this effect, recall that this suppression goes like 1-P, where P=(Pe-+Pe+)/(1-Pe-Pe+) is the effective polarization.
There should be a negligible contamination due to e+e-→e+e-γ. This assumption is detector and machine dependent and requires more work to be established. In principle it is possible to eliminate this background by requesting a photon with sufficient transverse momentum, which guarantees the appearance of an energetic e+/e- in the forward electromagnetic calorimeters. This demands perfect vetoing of electrons in these calorimeters which is only possible if the beam background remains at a manageable level. A careful optimization of the final focus region is needed to avoid overloading the very forward calorimeters (see for instance reference [13]).

### IV.2 Scenario 1

One has:

$$\sigma v = |g_V^X|^2 K^2 \sum_f n_{cf} |g_A^f|^2 \frac{2m_X^2 + s}{12\pi\left[(s-m_{Z'}^2)^2 + (m_{Z'}\Gamma_{Z'})^2\right]}$$

This formula shows that the annihilation cross section is independent of $\langle v^2 \rangle$.

A heavy Z' requires a large coupling to DM and therefore a wide resonance decaying *mostly invisible*. This type of scenario has already been discussed in [10]. If the Z' couplings to standard fermions are not suppressed with respect to the SM, the limits set by ATLAS/CMS for lepton pairs are still able to exclude this solution. Assuming a reduction factor K²~0.1 on the standard couplings allows to reach a wide domain of solutions as discussed below.

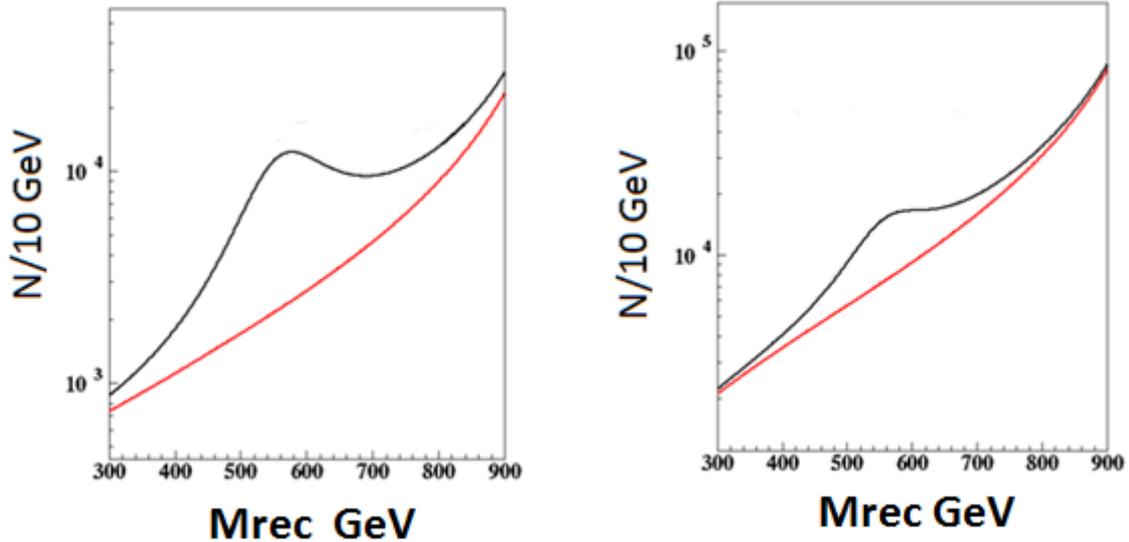

Figure 2a: Number of expected ISR events in 10 GeV bins versus the effective center of mass energy. The red curve shows the expected background. The black curve is the predicted rate assuming a Z' with $m_{Z'}$=550 GeV. These curves correspond to an ILC operated at 1 TeV and collecting 1 ab-1 with beam polarizations Pe-=0.9 and Pe+=-0.6. Figure 2b: Same as figure 2a with Pe-=0.8 and Pe+=-0.3.

Since one is dealing with Dirac fermions, the total width reads:

$$\Gamma(Z' \to X\bar{X}) = \frac{|g_V^X|^2 \, v m_{Z'}}{12\pi}$$

As already mentioned, one can use an ISR method requesting a photon emitted inside the detector. Measuring its energy k, one can determine the recoil mass from the expression:

$Mrec = ECM \sqrt{1-x}$ where ECM is the center of mass energy and x=2k/ECM. One can assume $m_{Z'}$=550 GeV and an ILC operating at ECM=1 TeV at full luminosity (1 ab-1) and with improved polarization, as previously defined. Figure 2 shows that, as expected, Mrec peaks at the Z' mass. The fast rising background is due to W exchange. The significance of the signal is very high since at the resonance one counts about 10000 events per 10 GeV bin with an expected background of ~2000 events. Figure 2b assumes more conservative assumptions on beam polarization (Pe-=80% and Pe+=-30%) and shows that a signal excess is still observable.

It should be underlined that this invisible Z' scenario is uniquely covered with this ISR method at ILC and could escape to direct observation into lepton pairs at LHC.

From figure 2, one concludes that this method allows to measure the Z' mass, its total width $\Gamma_t \sim \Gamma_{inv}$ and its invisible cross section at resonance σ~BR$_{ee}$BR$_{inv}$ with BR$_{inv}$~1. From these two measurements, one extracts the couplings of Z' to DM and to e+e-, which allows to draw some important clues about the underlying model.

How are these conclusions affected when one varies $m_{Z'}$ and $m_X$ ? Figure 3a and 3b indicate the allowed mass domains for $m_{Z'}$ versus $m_X$. It was checked that ILC operating at 1 TeV and collecting 1 ab-1 can fully cover the white allowed area.

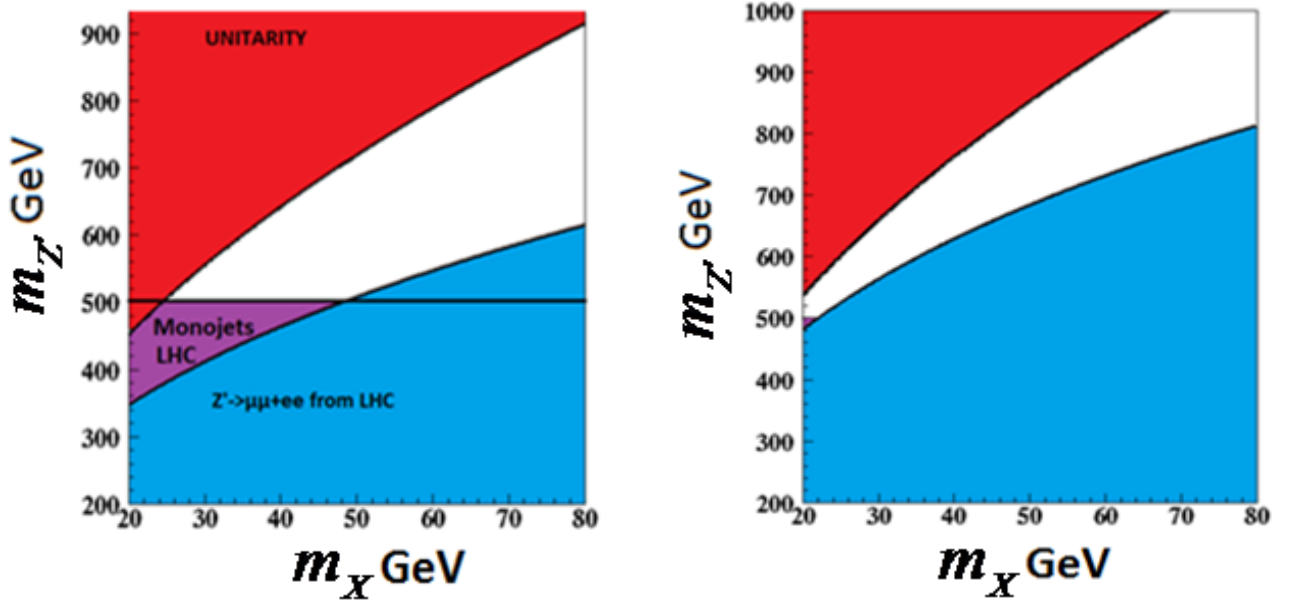

Figure 3a: corresponds to scenario 1 with K²=0.1 The red region corresponds to a Z' coupling to DM beyond the unitarity limit while the blue domain is excluded by present searches of Z' into lepton pairs. The purple domain would correspond to a visible excess of mono-jets plus missing energy. Figure 3b : Same as 3a with K²=0.2.

### IV.2 Scenario 2

In this scenario one has:

$$\sigma v = |g_A^X|^2 K^2 \sum_f n_{cf} (|g_V^f|^2 + |g_A^f|^2) \frac{s\langle v^2 \rangle}{12\pi \left[(s-m_{Z'}^2)^2 + (m_{Z'}\Gamma_{Z'})^2\right]}$$

where the suppression factor K² is assumed uniform and ~0.1.

Since one is dealing with Majorana fermions, the total width reads:

$$\Gamma(Z' \to X\bar{X}) = \frac{|g_A^X|^2 v^3 m_{Z'}}{24\pi}$$

The next steps are very similar to the previous scenario with similar conclusions.

### IV.3 Comparison with LHC

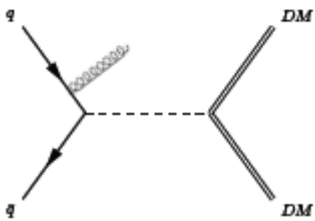

A quasi-invisible Z' could, in principle, also be observed at LHC using the mono-jet technique, as Illustrated by above diagram which is of course reminiscent of the ISR technique in e+e- . To compare the two types of colliders, one notes that, at LHC, the invisible Z' cross section goes like K²g²$_{Zqq}$BR$_{Z'inv}$ with BR$_{Z'inv}$~1. Since BR$_{Zinv}$=0.2, the rate is reduced by 5K²RL with respect to the inclusive production of invisible Z where RL is a luminosity ratio which takes care of the difference of mass between Z and Z'. This ratio tends to 1 when $m_{Z'}$<<Etmiss, where Etmiss is the transverse energy carried by the gluon. With the most recent data [14,15] taken at 8 TeV, the sensitivity of LHC is reaching Etmiss~800 GeV. For $m_{Z'}$= 500GeV and Etmiss=800 GeV, one has RL~0.6. The expected excess over the Z contribution would be ~+30%, barely observable. LHC at 14 TeV will allow to reach $m_{Z'}$=1 TeV. Clearly these conclusions may be moved around by modifying the parameter K².

Figure 3 summarizes, fixing K²=0.1 the mass regions already excluded by present searches of Z' into lepton pairs (blue) and indicates the region sensitive to mono-jets. The red area is excluded by the unitarity limit, $g_A^{X2}$ <4π. This figure indicated that for $m_X$~35 GeV, LHC will soon cover this scenario.

If LHC observes only an excess at large Etmiss and no lepton pair signal, the interpretation of such a signal would be uncertain since, without the initial state energy-momentum constraint, one cannot observe the resonance shape shown in figure 2.

In conclusion, for the *'invisible Z' '* scenario, ILC provides a unique opportunity of detection based on radiative return and background suppression using highly polarized beams. Within the Fermi-LAT scenario, there are good prospects of discovery at LHC for this scenario.

## V SM Higgs portal

Higgs couplings can be written as:

$$\mathcal{L}_{\text{int}} \supset \left[ \bar{X} \left( \lambda_s^X + \lambda_p^X i\gamma^5 \right) X + \bar{f} \left( \lambda_s^f + \lambda_p^f i\gamma^5 \right) f \right] A$$

As explained in section II, one only retains the $\bar{X}\gamma_5 X$ couplings. For a SM Higgs $\lambda_s^b$ =0.013 and $\lambda_p^b$ =0. Then one has:

$$\sigma v = \frac{nc |\lambda_s^X \lambda_p^X|^2 s}{16\pi(s-m_h^2)^2}$$ where s=4$m_X$²

For $m_X$=35 GeV, one has $\lambda_s^b \lambda_p^X$ =3.1% and hence $\lambda_p^X$ =2.4

What would be the invisible h width? For a Majorana X:

$$\Gamma(h \to XX) = \frac{|\lambda_p^X|^2 v^3 m_h}{16\pi}$$ Therefore $\Gamma_{\text{inv}}$=8 GeV, to be compared to the total width which is

about 4 MeV. This is of course completely excluded by present LHC constraints [16].

→ Therefore the SM Higgs portal cannot explain the Fermi-LAT effect.

What happens if one increases the DM mass? One finds that that Higgs coupling to DM vanishes rapidly. For $m_X$ above 60 GeV the invisible branching goes below 10% and therefore can only be excluded at e+e- colliders by using the Zh mode, with Z decaying into lepton pairs, which allows a precise recoil mass reconstruction. In this way, e+e- machines can provide a model independent measurement of the invisible Higgs width and one can measure the invisible branching ratio down to a % level [17].

If $m_X$>$m_h$/2 one can still produce a virtual h*, but the cross section decreases rapidly and the increase in mass coverage is marginal as reported in reference [18]. Reference [19] has envisaged the possibility to use the fusion process e+e→ZZe+e- with ZZ→h*. When h* decays invisibly, it is still possible to reduce the backgrounds by using the final state leptons. At 3 TeV center of mass energy, reachable by CLIC, the increase in mass coverage is also marginal with the predicted Higgs DM couplings.

## VI Gauge singlet DM

This model is truly a minimal extension of the SM since it only adds a stable scalar singlet which only couples to the Higgs boson and plays the role of DM. The present limit on Higgs invisible decays forbids that the mass of this singlet falls below ~53 GeV and therefore this mechanism is marginally compatible with the Fermi-LAT signal. Improving on the invisible Higgs decay width limit at e+e- colliders will only allow to reach a mass limit close to ~56 GeV.
For masses above mh/2, reference [20] predicts a direct signal with no collider counterpart.

## VII Higgs singlets

In this model one assumes that fermionic DM can annihilate through BSM Higgs gauge singlets, scalar and/or axial, called s and a. Contrary to previous section, these particles can also couple to ordinary fermions.

To estimate the size of the couplings of s/a to SM and DM fermions, we need to introduce the DM constraints. One has the following expression:

$$<\sigma v> = \frac{3|\lambda_a^b \lambda_a^X|^2 s}{16\pi(s-m_a^2)^2} + \frac{3|\lambda_s^b \lambda_s^X|^2 \langle v^2 \rangle s}{16\pi(s-m_s^2)^2} + \frac{3\left(|\lambda_s^b \lambda_a^X|^2 + |\lambda_a^b \lambda_s^X|^2 \langle v^2 \rangle\right)s}{16\pi(s-m_s^2)(s-m_a^2)}$$

where s=4$m_X^2$ and where one assumes that the largest coupling is due to the quark b by analogy to the Higgs and to account for the GC signal. To avoid the <v>=0.3 suppression effect and forgetting about the Sommerfeld enhancement, one is led to assume that the first term dominates which naturally occurs if $m_a$~$m_s$.

As reported in [1], direct detection limits are not sensitive to this type of couplings given that there is a large momentum transfer suppression.

At LHC the top loops allow production of a/s and their decay into 2γ. The visibility of this signal cannot be predicted since, from DM constraint, one can only estimate the product $\lambda_a^b \lambda_a^X$. For instance, for ma=50 GeV, one predicts $\lambda_a^b \lambda_a^X$ ~0.007. Taking tentatively the b coupling similar to SM, ~1%, the coupling to X will be dominant ~0.5 meaning that 'a' will decay invisibly (>99%) precludes its direct detection.

In e+e-, s-channel production is severely suppressed and Higgstrahlung does not operate for gauge singlet unless they mix with h which, for a SM h, can only happen for s. If a/s are lighter than mh/2, h could decay into ss or aa, presumably invisible but detectable at a Higgs factory by using the recoil mass technique.

If there is h-s mixing, one could have other measurable consequences, the most obvious one being the hZZ coupling suppressed by cosα, where α defines the mixing angle between h and s. In e+e- this coupling is measured to better that 1% which corresponds to a mixing angle ~ 0.1. Since h carries a component s, it could decay invisibly providing an additional clue. Taking the numerical example previously given, one has BR(h→XX)/BR(h→bb)= $|\lambda_a^X|^2 \tan^2\alpha / |\lambda_s^b|^2$ , which appears promising since one expects that $\lambda_a^X / \lambda_s^b \gg 1$.

In summary, this scenario can be directly tested by searching for invisible decays of the SM Higgs mode. Indirect observation can be achieved by measuring a reduced hZZ coupling.

# VIII Higgs doublets

## VIII.1 The invisible A scenario

This scenario is based on the 2 doublet model and offers large freedom:

- The mass of A is free
- As a pseudo-scalar, A cannot mediate a SI DM-nucleon scattering, avoiding direct detection bounds
- The b coupling enhanced if tanβ>1

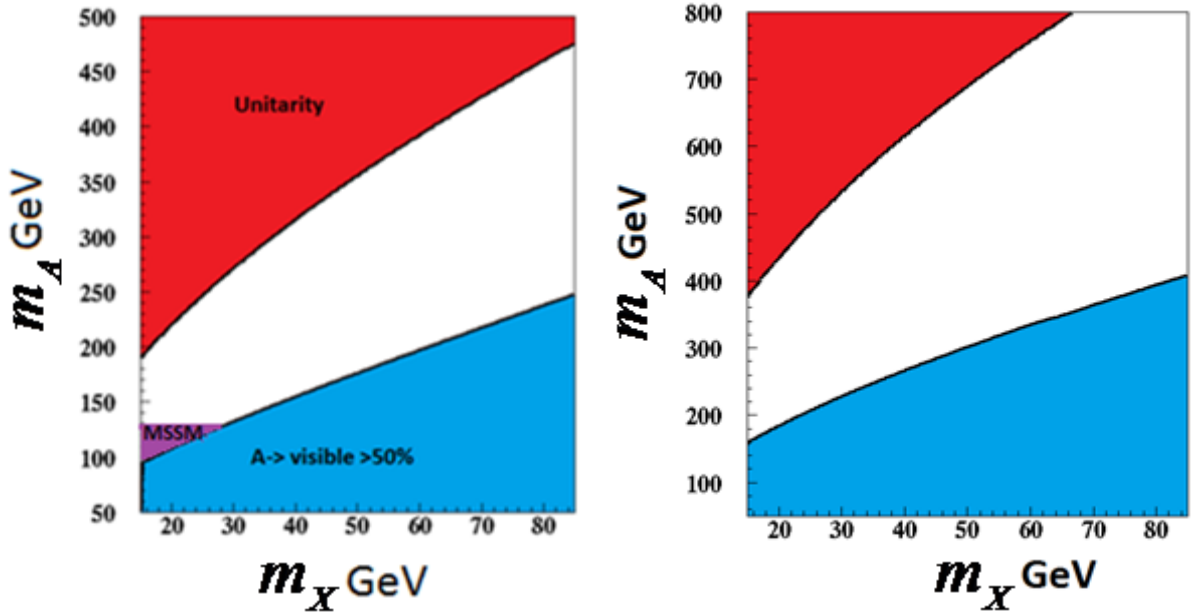

Figure 4a: For the A portal solution with tanβ=5 the blue area corresponds to A decaying >50% visibly, therefore observable at LHC. The red area corresponds to couplings to DM beyond the unitarity limit. Figure 4b: same with tanβ =20. The purple area is already excluded by LHC assuming MSSM.

Again one assumes an axial coupling of A to a Majorana X with no coupling to Z:

$$\sigma v = \frac{3|\lambda_A^b \lambda_A^X|^2 s}{16\pi(s-m_A^2)^2} \text{ with } \lambda_A^b = \frac{g m_b \tan\beta}{2 m_W} \quad \Gamma(A\to XX) = \frac{|\lambda_A^X|^2 v m_A}{16\pi} \text{ and } \Gamma(A\to b\bar{b}) = \frac{3|\lambda_A^b|^2 m_A}{8\pi}$$

Assuming $m_A$=300 GeV, tanβ=10 and $m_X$=35 GeV, one has $\lambda_A^b \lambda_A^X$ =0.25, hence $\lambda_A^X$ =2.

With this value, an on mass shell A decays visibly in ~2.5% of the cases. In principle A can also decay into Zh but, for a heavy A, the ZhA coupling is too small to contribute significantly.

While this solution requires an extended Higgs sector, it satisfies all present constraints. In particular LHC cannot exclude this solution given that A decays invisibly in >90% of the cases. For what concerns H, if heavy enough, its main decay would be into hh. This mode has been searched at LHC, using h decays into two photons and 4 leptons. In [21,22] one finds that this search applies only for tanβ~1.

With such a scenario one expects $m_H$~$m_A$. Figure 4 displays the mass domain expected for this type of solution. The channel HA would be accessible to a TeV e+e- collider provided that $m_A$<500 GeV. It would allow to tag the presence of invisible decays of A by using a recoil mass technique by reconstructing the accompanying H boson. Typically, for an integrated luminosity of 1 ab-1, one expects ~7 000 HA events [23] with A decaying mostly invisibly.

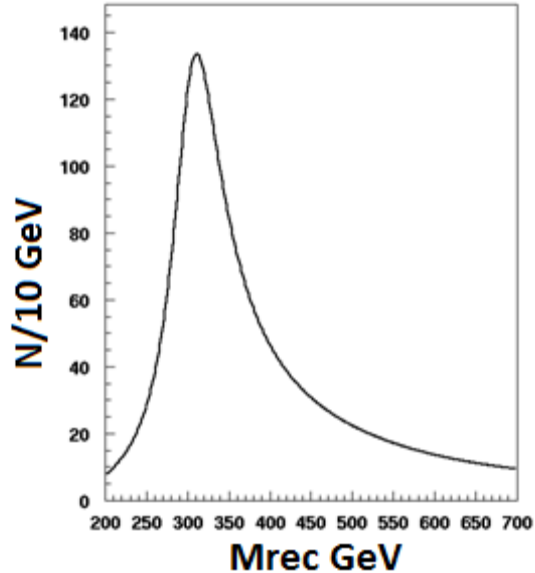

Figure 5: For the A portal solution with tanβ=5 and $m_A$=300 GeV, the predicted recoil mass distribution using the H decay into 4b. The bump corresponds to A decaying invisibly in the HA final state for ECM=1 TeV and with a luminosity of 1 ab-1.

For what concerns H, which will serve to tag the presence of an invisible A, the standard decay mode is dominated by hh (98%). One has the freedom to assume that H does not couple to DM and therefore it should be straightforward to reconstruct H decays which can be distinguished from the SM background. ILC detectors are optimized to perform this type of analysis with precise jet energy measurement (3% resolution). One can use heavy quark identification for h decaying predominantly into pairs of b quarks. The main background, which comes from top pairs producing only 2 b jets, is easily rejected.

Figure 5 shows the expected recoil mass distribution obtained using, for the HA channel at ECM=1 TeV, a realistic energy resolution for the H decaying into 4b (from the 2h) and including initial state radiation. The A resonance parameters can be precisely measured with $m_A$=300±0.8 GeV and $\Gamma_A$ =24±1 GeV. From the later one can extract the coupling $\lambda_A^X$ with 2% accuracy.

It is also possible to assume that H also couples axially to DM. The large coupling of H to 2h implies that the visible fraction of H decays will be large, ~50%, meaning that the recoil mass technique previously described can also be applied for this hypothesis.

### VIII.2 The invisible A scenario at LHC

Recall that at LHC standard searches [24] based on H/A single production do not operate for tanβ~5. If A decays invisibly one can still keep some coverage using H decays alone. So far the most constraining limits come from the mass limit [25] on H$^\pm$ and the MSSM relation m²H$^\pm$=m²W$^\pm$+m²A which implies that mA>125 GeV. This relation can however be relaxed with NMSSM. As can be seen from figure 4, the GC excess solution with =35 GeV corresponds to mA >150 GeV, not excluded by LHC.

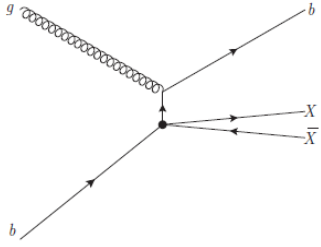

Reference [26] suggests using the mechanism sketched in above diagram: a gluon scatters a b quark from the sea which radiates a A boson decaying into DM. The mono-jet in this case is a b-jet which allows to tag this mechanism. Reference [3] indicates that the required sensitivity is sill way below what is needed to observe this signal. This sensitivity depends on the coupling of A to b quarks which is proportional to tanβ .

In reference [27], the present scenario has also been considered and, similarly to the invisible Z' case, one can apply the mono-jet technique. While the present sensitivity does not allow to set a meaningful limit, [27] predicts that with 14 TeV and 40 fb-1 integrated luminosity, it would become possible to cover this type of scenario.

Note finally that these conclusions substantially differ from those of [28] which study an NMSSM scenario with resonant coupling of DM to a very light A. The main reason for this difference comes from our freedom to assume a large (although bounded by the unitarity limit) coupling constant of A to DM while there are restrictions within NMMSM for the couplings to neutralinos.

While these prospects appear promising, it will be difficult to interpret unambiguously the origin of an excess of mono-jet production, for instance as due to A or to a Z'. One may of course hope that other signals due to a 2 doublet scenario will orient the interpretation.

## Conclusion

The DM candidate from Fermi-LAT [1] has been interpreted in terms of the 2 SM portals: annihilation through Z or H bosons and through 2 BSM portals: annihilation through Z' or A bosons. Prospects for DM discovery at e+e- colliders were presented and appear promising.

To cope with the DD limit, one is led to assume an *axial coupling of DM* to the Z boson which naturally enhances the coupling to b quarks and comforts the interpretation of [1]. One however finds an inconsistent picture for the rate of annihilation at present temperatures, unless DM receives the strong *Sommerfeld enhancement* predicted by models which try to reconcile the DM distribution at galactic scales.

The *invisible Z width* is the most sensitive SM observable to monitor this scenario. With the accuracy given by LEP1, one can already disfavor $m_X$ <27 GeV. Future e+e- colliders will reach $m_X$ <35 GeV.

For what concerns the Z' portal, there is a similar scenario leading to the same inconsistent picture unless one advocates a Sommerfeld enhancement. There is however another scenario, where DM is a Dirac fermion and Z' has purely axial couplings to SM fermions. This scenario *satisfies all constraints*:

- It is far from being excluded by direct searches
- It allows to get a consistent picture without invoking a Sommerfeld enhancement.

In both scenarios, due to LHC limits on di-lepton searches, one needs to assume reduced couplings to SM fermions, therefore a *quasi-invisible Z' decay*. This Z' is accessible to a TeV e+e- collider trough radiative return. This technique allows to observe the Z' resonance and determine its mass, width and coupling to e+e-.

Generally speaking, the ISR technique in e+e- provides a powerful tool to detect DM, provided that one can run this collider with highly polarized beams to eliminate the e+e- →$\nu_e\nu_e\gamma$ process due to W exchange. It also requires an optimized set up to fully eliminate the contamination from e+e-→e+e-$\gamma$.

A SM Higgs portal interpretation implies a large Higgs invisible width, already excluded by LHC results. Non-minimal scenarios with extra Higgs doublets or singlets provide promising scenarios.

A *singlet Higgs scalar boson* could mix with h and modify SM couplings and generate an invisible width observable in e+e-. If the singlet Higgs boson is light, the SM Higgs could decay into ss or aa, s and a being the scalar and axial singlet bosons. These modes produced associated to a Z are observable in e+e- down to a high sensitivity.

In a *two doublet extension*, one can assume that DM annihilation proceeds through an axial boson A. This scenario naturally leads to the b jet interpretation and does not require a Sommerfeld enhancement. It also naturally avoids LUX constraints. The pseudo-scalar *Higgs boson A decaying invisibly* can be observed in e+e- when produced in association to a H boson in a TeV e+e- collider. One can precisely measure its mass and width.

In summary, unless one invokes a strong Sommerfeld enhancement for DM annihilation at present energies, the overall conclusion is that a BSM mediator, for instance a *Z' or extra Higgs bosons*, is needed to interpret the Fermi-LAT DM evidence. If true, these interpretations predict scenarios which could already be tested at LHC while a TeV e+e- collider should provide an essential tool for a precise measurement of the parameters of this resonance.

**Acknowledgements:** G. Arcadi and Y. Mambrini acknowledge support from the ERC advanced grant Higgs@LHC.